\documentclass[11pt]{article}
\usepackage{amsfonts,amssymb, graphicx,a4}

\begin{document}


\voffset=-1.5truecm\hsize=16.5truecm    \vsize=24.truecm
\baselineskip=14pt plus0.1pt minus0.1pt \parindent=12pt
\lineskip=4pt\lineskiplimit=0.1pt      \parskip=0.1pt plus1pt

\def\ds{\displaystyle}\def\st{\scriptstyle}\def\sst{\scriptscriptstyle}


\global\newcount\numsec\global\newcount\numfor
\gdef\profonditastruttura{\dp\strutbox}
\def\senondefinito#1{\expandafter\ifx\csname#1\endcsname\relax}
\def\SIA #1,#2,#3 {\senondefinito{#1#2}
\expandafter\xdef\csname #1#2\endcsname{#3} \else
\write16{???? il simbolo #2 e' gia' stato definito !!!!} \fi}
\def\etichetta(#1){(\veroparagrafo.\veraformula)
\SIA e,#1,(\veroparagrafo.\veraformula)
 \global\advance\numfor by 1
 \write16{ EQ \equ(#1) ha simbolo #1 }}
\def\etichettaa(#1){(A\veroparagrafo.\veraformula)
 \SIA e,#1,(A\veroparagrafo.\veraformula)
 \global\advance\numfor by 1\write16{ EQ \equ(#1) ha simbolo #1 }}
\def\BOZZA{\def\alato(##1){
 {\vtop to \profonditastruttura{\baselineskip
 \profonditastruttura\vss
 \rlap{\kern-\hsize\kern-1.2truecm{$\scriptstyle##1$}}}}}}
\def\alato(#1){}
\def\veroparagrafo{\number\numsec}\def\veraformula{\number\numfor}
\def\Eq(#1){\eqno{\etichetta(#1)\alato(#1)}}
\def\eq(#1){\etichetta(#1)\alato(#1)}
\def\Eqa(#1){\eqno{\etichettaa(#1)\alato(#1)}}
\def\eqa(#1){\etichettaa(#1)\alato(#1)}
\def\equ(#1){\senondefinito{e#1}$\clubsuit$#1\else\csname e#1\endcsname\fi}
\let\EQ=\Eq


\def\bb{\hbox{\vrule height0.4pt width0.4pt depth0.pt}}\newdimen\u
\def\pp #1 #2 {\rlap{\kern#1\u\raise#2\u\bb}}
\def\hhh{\rlap{\hbox{{\vrule height1.cm width0.pt depth1.cm}}}}
\def\ins #1 #2 #3 {\rlap{\kern#1\u\raise#2\u\hbox{$#3$}}}
\def\alt#1#2{\rlap{\hbox{{\vrule height#1truecm width0.pt depth#2truecm}}}}

\def\pallina{{\kern-0.4mm\raise-0.02cm\hbox{$\scriptscriptstyle\bullet$}}}
\def\palla{{\kern-0.6mm\raise-0.04cm\hbox{$\scriptstyle\bullet$}}}
\def\pallona{{\kern-0.7mm\raise-0.06cm\hbox{$\displaystyle\bullet$}}}


\def\data{\number\day/\ifcase\month\or gennaio \or febbraio \or marzo \or
aprile \or maggio \or giugno \or luglio \or agosto \or settembre
\or ottobre \or novembre \or dicembre \fi/\number\year}

\setbox200\hbox{$\scriptscriptstyle \data $}

\newcount\pgn \pgn=1
\def\foglio{\number\numsec:\number\pgn
\global\advance\pgn by 1}
\def\foglioa{a\number\numsec:\number\pgn
\global\advance\pgn by 1}



\def\sqr#1#2{{\vcenter{\vbox{\hrule height.#2pt
\hbox{\vrule width.#2pt height#1pt \kern#1pt
\vrule width.#2pt}\hrule height.#2pt}}}}

\let\ciao=\bye \def\fiat{{}}
\def\pagina{{\vfill\eject}} \def\ni{\noindent}
\def\bra#1{{\langle#1|}} \def\ket#1{{|#1\rangle}}
\def\media#1{{\langle#1\rangle}} \def\ie{\hbox{\it i.e.\ }}
\def\i{\infty}

\def\dpr{\partial} 
\def\V#1{\vec#1} \def\Dp{\V\dpr}
\def\oo{{\V\o}} \def\OO{{\V\O}} \def\uu{{\V\y}} \def\xxi{{\V \xi}}
\def\xx{{\V x}} \def\yy{{\V y}} \def\kk{{\V k}} \def\zz{{\V z}}
\def\rr{{\V r}} \def\zz{{\V z}} \def\ww{{\V w}}
\def\Fi{{\V \phi}}
\def\I{{\rm I}}
\def\Rar{\Rightarrow}
\def\rar{\rightarrow}
\let\LRar=\Longrightarrow

\def\lh{\hat\l} \def\vh{\hat v}

\def\ul#1{\underline#1}
\def\ol#1{\overline#1}

\def\ps#1#2{\psi^{#1}_{#2}} \def\pst#1#2{\tilde\psi^{#1}_{#2}}
\def\pb{\bar\psi} \def\pt{\tilde\psi}

\def\E#1{{\cal E}_{(#1)}} \def\ET#1{{\cal E}^T_{(#1)}}
\def\LL{{\cal L}}
\def\RR{{\bf R}}
\def\SS{{\cal S}} \def\NN{{\cal N}}
\def\HH{{\cal H}}\def\GG{{\cal G}}\def\PP{{\cal P}} \def\AA{{\cal A}}
\def\BB{{\cal B}}\def\FF{{\cal F}}

\def\tende#1{\vtop{\ialign{##\crcr\rightarrowfill\crcr
              \noalign{\kern-1pt\nointerlineskip}
              \hskip3.pt${\scriptstyle #1}$\hskip3.pt\crcr}}}
\def\otto{{\kern-1.truept\leftarrow\kern-5.truept\to\kern-1.truept}}
\def\arm{{}}
\font\bigfnt=cmbx10 scaled\magstep1
\def\GI{\mathbb{G}}
\def\VU{\mathbb{V}}
\def\ED{\mathbb{E}}

\def\a{\alfa}
\def\b{\beta}
\def\d{\delta}
\def\e{\varepsilon}
\def\f{\varphi}
\def\g{\gamma}
\def\l{\lambda}
\def\m{\mu}
\def\v{\vu}
\def\p{\pi}
\def\ph{\varphi}
\def\s{\sigma}
\def\r{\rho}
\def\t{\tau}
\def\D{\Delta}
\def\X{\Xi}
\def\G{\Gamma}
\def\La{\Lambda}


\begin{center}
{\Large Potts model  on  
infinite graphs 
and the limit of chromatic polynomials}
\vskip.4cm
{\bf  Aldo Procacci$^*$\footnote{Partially supported by CNPq (Brazil)},~
Benedetto Scoppola$^\dag$\footnote{ Partially supported by CNR, G.N.F.M. (Italy)} ~ and 
Victor Gerasimov$^*$}

\end{center}
\vskip1.0cm
\begin{center}
{\small  $^*$Departamento de  Matem\'atica -  Universidade Federal de Minas Gerais

Av. Ant\^onio Carlos, 6627 - Caixa Postal 702 - 30161-970 - Belo Horizonte - MG 
Brazil

e-mails: aldo@mat.ufmg.br (Aldo Procacci); victor@mat.ufmg.br (Victor Gerasimov)
              
\vskip.2cm
$^\dag$Dipartimento di Matematica -  Universit\'a ``La Sapienza'' di Roma

Piazzale A. Moro 2, 00185 Roma, Italy 

e-mail: benedetto.scoppola@roma1.infn.it

~

~
}
\end{center}

\begin{abstract}
\ni {\it 
Given  an infinite   graph $\GI$ quasi-transitive and amenable
with maximum degree $\D$,
we show that  reduced ground state degeneracy per site $W_r(\GI,q)$ of  the $q$-state
antiferromagnetic   Potts model at zero temperature 
on $\GI$ is analytic in the variable $1/q$,  whenever $|2\D e^3/q|< 1$.
This result proves, in an even
stronger formulation,  a conjecture
originally sketched   in \cite{KE} and explicitly formulated in
\cite{ST}, based on which a sufficient condition for 
$W_r(\GI,q)$ to  be analytic at $1/q=0$ is that $\GI$ is a regular lattice.
}
\end{abstract}

\vskip.5cm
\ni {\small {\bf Keywords}: {\it Potts model, chromatic polynomials, cluster expansion}}

\vskip.5cm
\ni\S 1 {\bf Introduction}
\vskip.3cm
\numsec=1\numfor=1
\ni The Potts model with $q$ states (or $q$ ``colors'') is a system
of random variables (spins) $\sigma_x$
sitting in the vertices $x\in \VU$
of a locally finite graph $\mathbb{G}=(\mathbb{V}, \mathbb{E})$ with vertex set
$\VU$ and edge set $\mathbb{E}$, and
taking values  in the set of integers 
$\{1,2, \dots, q\}$ .  Usually the graph $\GI$ is a regular
lattice, such as $\mathbb{Z}^d$ with the set of edges $\mathbb{E}$ being the
set on nearest neighbor pairs, but of course more general situations can be
considered.
A {\it configuration} $\sigma_\VU$ of the system is
a function $\sigma_\VU: \VU\to \{1,2, \dots, q\}$ with
$\sigma_x$ representing the value of the {\it spin} at the site $x$. 
We denote by $\G_{\mathbb{V}}$ the
set of all spin configurations in $\mathbb{V}$. If $V\subset\mathbb{V}$ 
we denote  $\sigma_{V}$ the restriction of $\sigma_{\mathbb{V}}$ to  $V$ and by
$\G_{V}$ the set of all spin configurations in $V$.

 Let $V\subset \VU$ and  let $\GI|_{V}=(V, \ED_V)$,
where $\mathbb{E}|_V=\{\{x,y\}\in \mathbb{E}: x\in V, y\in V\}$. Then for $V\subset \VU$ {\it finite},
the {\it energy of the spin configuration $\sigma_V$ in  $\GI|_{V}$} is defined as
$$
H_{\GI|_{V}}(\sigma_{V}) =
-J\sum_{\{x,y\}\subset \mathbb{E}|_V}\d_{\sigma_x \sigma_y}  \Eq(2.1.0)
$$
where
$\d_{\sigma_x\sigma_y}$ is the Kronecker symbol which is equal to one when $\sigma_x=\sigma_y$ and
zero otherwise. 
The {\it coupling}  $J$ is called
{\it ferromagnetic} if $J> 0$ and {\it anti-ferromagnetic} if
$J<0$. 

 The {\it statistical mechanics} of the system can be done by
introducing the {\it Boltzmann weight} of a configuration $\sigma_V$, defined as
$\exp\{-\b H_{\GI|_{V}}(\sigma_{V})\}$ where $\b\ge 0$ is the inverse temperature. Then the
{\it probability} to find the system in the configuration $\sigma_V$ is given by
$$
{\rm Prob}(\sigma_V)= {e^{-\b H_{\GI|_{V}}(\sigma_{V})}\over
Z_{\GI|_{V}}(q)}\Eq(1.2.0)
$$
\ni The normalization constant in the denominator 
is called the  {\it partition function} and is given by
$$
Z_{\GI|_{V}}(q,\b)=\sum_{\sigma_{V}\in \G_V} e ^{- \b H_{\GI|_{V}}(\sigma_{V})} \Eq(2.4b.0)
$$
The case $\b J=-\i$ is the {\it anti-ferromagnetic} and 
{\it zero temperature} Potts model with $q$-states. In this case 
configurations with non zero probability are only those in 
which adjacent spins have different values (or colors) and 
$Z_{\GI|_{V}}(q)$ becomes simply the number of all allowed 
configurations.
 
 The {\it thermodynamics} of the system at inverse temperature $\b$ and
``volume'' $V$ is recovered through the {\it free energy per unit volume} given
by
$$
f_{\GI|_{V}}(q,\b)={1\over |V|} \ln Z_{\GI|_{V}}(q)\Eq(free)
$$
where $|V|$ denotes the cardinality of $V$. All thermodynamic functions of the
system can be obtained via derivative of the free energy. In the zero temperature
anti-ferromagnetic case the function $f_{\GI|_{V}}(q,\b)$ is usually called 
{\it the ground state entropy} of the system.

The Potts model, despite its simple formulation,
is a intensely  investigated subject. 
Besides its own interest as a statistical mechanics model,
it has deep connections with several areas in theoretical
physics, probability and combinatorics. 

In particular, Potts models on  general graphs are strictly  related to a typical
combinatorial problem.
As a matter of fact, the partition
function of the  Potts model with $q$ state
on a finite graph $G$, is equal, in the zero temperature
antiferromagnetic case,
to the number of proper 
coloring with $q$ colors of the graph $G$,
where proper coloring means that adjacent vertices of the graph must have
different colors. This number viewed as a function of
the number of colors $q$ is actually  a polynomial function in the variable $q$
which is known as {\it chromatic polynomial}. On the other hand,
the same partition function in the general case can be related to 
more general chromatic type polynomials, known as {\it Tutte polynomials} \cite{Tu}.
This beautiful connection between
statistical mechanics and graph coloring problems, 
first discussed by  Fortuin and Kasteleyn
\cite{FK}, has been extensively studied
and continues to attract many researchers till nowadays (see
e.g. \cite{Bax}, \cite{CS}, \cite{SS}, \cite{Sh}, \cite{So}, \cite{So2}, \cite{WM}
and reference therein).

One of the main 
interests in statistical physics is to establish whether or not a given system
exhibits {\it phase transitions}. 
This means  to search for points in the 
interval $\b\in [0,\i]$  where some thermodynamic function (like 
e.g. the free energy defined above) is  non analytic. 
Now, functions  as \equ(2.4b.0) and \equ(free) are manifestly analytic as long
as $V$ is  a finite set. Hence phase transition (i.e. non-analyticity)
can arise only in the so called {\it infinite volume limit} or {\it thermodynamic
limit}. 
That is, the graph $\GI$
is some countably infinite graph, usually a regular lattice, and the
infinite volume limit 
$$
f_{\GI}(q,\b)=\lim_{N\to \i}{1\over |V_N|} \ln Z_{\GI|_{V_N}}(q,\b)\Eq(freinf)
$$
is taken along a sequence $V_N$ of finite subsets
of $\VU$ such that, roughly speaking, $\GI|_{V_N}$ increase in size equally in
all directions. Tipically, when $\mathbb{V}$ is $\mathbb{Z}^d$, $V_N$ are usually cubes
of increasing size $L_N$ .
There is  a considerable amount of rigorous results about thermodynamic limit
and phase transitions 
for the Potts model on $ \mathbb{Z}^d$ and other regular lattices, see e.g. the reviews \cite{Y}
and, more recently, \cite{So2}.

 On the other hand, the study of thermodynamic limits of spin systems 
on infinite graphs which are not usual lattices 
has recently  drove the attention of many researcher (e.g. 
\cite{BS},  \cite{HSS}, \cite{J}, 
\cite{L} and references therein).

Concerning specifically the antiferromagnetic Potts model and/or chromatic 
polynomial on infinite graphs, the problem of the thermodynamic limit
was first considered  by Biggs \cite{Bi} with further discussions in \cite{BM}
and in \cite{KE}.
 Very recently  Sokal \cite{So} has  shown that
for any {\it finite graph} $G$ with maximum degree $\D$, the zeros of the
chromatic polynomial lies in a disk $q\le C\D$ where $C$ is a constant.
An important extension of  this result
would be to prove the existence and analyticity of the limiting
free energy per unit volume \equ(freinf) for a suitable class, as wide as possible,
of infinite graphs. Such a generalization would be relevant from the
statistical mechanics point of view, since it would imply that 
anti-ferromagnetic Potts model on
such class of infinite graphs, if $q$
is sufficiently large,  does not present a phase  transition at zero
temperature (and hence at any temperature).

 To this respect, Shrock and Tsai  have explicitly formulated a conjecture \cite{ST}
(see also \cite{KE}), based on which
the ground state entropy per unit volume of the antiferromagnetic Potts model at
zero temperature on an infinite graph $\GI$ should be analytic in the neighbor of  $1/q=0$ 
whenever $\GI$ is a regular lattice.

 In this paper 
we actually prove  that this conjecture is true not only for 
regular lattices, but even for
a wide calls of graphs. In particular  we prove that the 
ground state 
zero entropy is analytic near $1/q=0$ for all infinite graphs which are quasi transitive
and  amenable, and the limit may be evaluated along {\it any} F\o lner sequence
in $\VU$. 
We stress that this result 
proves  the Schrock  conjecture in a considerably
stronger formulation, since
{\it all regular 
lattices}, either with the elementary cell made  by one single vertex
or  by more than one vertex, are indeed
quasi-transitive amenable graphs
but actually the class of quasi-transitive amenable graphs is
much wider than that of regular lattices.

The paper is organized as follows. In section 2 we introduce the notations used along the paper, 
and we enunciate the main result  (theorem 2). In section 3
we rephrase the problem in term of polymer expansion  and prove
a main  technical result (lemma 4). In section 4 we prove a  graph theory property (lemma
6) concerning
quasi-transitive amenable graphs. Finally in section 5 we give the proof of the
main result of the paper, i.e. theorem 2.

\vskip.7cm
\ni\S2. {\bf Some further notations and statement of the main result}
\numsec=2\numfor=1
\vskip.3cm

\ni In general, if $V$ is any finite
set, we denote by $|V|$ the number of elements of $V$. The set
$\{1,2,\dots ,n\}$ will be denoted shortly $\I_n$.
We denote 
by ${\rm P}_2(V)$ the set of all subsets $U\subset V$ such
that $|U|=2$ and by ${\rm P}_{\ge 2}(V)$ the set of all {\it finite}
subsets $U\subset V$ such
that $|U|\ge 2$.

 Given a countable set $V$, and given $E\subset {\rm P}_2(V)$, the pair
$G=(V, E)$ is called  a {\it graph} in $V$. The elements of $V$
are called {\it vertices} of $G$ and the elements of $E$ are called
 {\it edges} of $G$. 
Given two graphs $G=(V,E)$ and $G'=(V',E')$ in $V$, we say that
$G'\subset G$ if  $E'\subset E$ and $V'\subset V$.

 Given a graph $G=(V, E)$, two vertices $x$ and $y$ in $V$ are said to be {\it adjacent}
if $\{x, y\}\in E$.
The {\it degree} $d_x$ of a vertex $x\in V$
in $G$ is the number of vertices $y$ adjacent to $x$.  A graph $G=(V,E)$ is said
{\it locally finite}
if $d_x<+\i$ for all $x\in V$, and it is said {\it bounded degree} if 
$\max_{x\in V} \{d_x\}\le \D<\i$. 
A graph $G=(V, E)$ 
is said to be {\it connected}
if for any pair $B, C$ of  subsets of $V$ such that
$B\cup C =V$ and $B\cap C =\emptyset$, there is an edge  $e\in E$ such
that $e\cap B\neq\emptyset$ and $e\cap C\neq\emptyset$.

\def\sg {{\rm supp} \,g}
\def\sG {{\rm supp} \,G}

 We denote by ${ \cal G}_{V}$ the set of all connected graphs
with vertex set $V$. If $V=\I_n$ we use the notation ${ \cal G}_{n}$ in place of
${ \cal G}_{\I_n}$.
A {\it tree} graph $\t$ on $V$
is a connected graph $\t\in {\cal G}_V$ such that $|\t|=|V|-1$.
We denote by ${ \cal T}_{V}$ the set of all  tree graphs
of $V$ and shortly  ${ \cal T}_{n}$ in place of
${ \cal T}_{\I_n}$.

 Let ${\bf R}_n\equiv(R_1 ,\dots ,R_n)$ an ordered n-ple of non empty sets, then 
we denote by $E(\RR_n)$ the set $\subset {\rm P}_2(\I_n)$ defined
as $E(\RR_n)=\{\{i,j \}\in {\rm P}_2(I_n): \, R_i\cap Rj\neq \emptyset\}$. We denote
$G(\RR_n)$ the graph $(\I_n, E(\RR_n))$.

 Given two distinct vertices $x$ and $y$ of $G=(V,E)$, a {\it path} $\t(x,y)$
joining $x$ to $y$  is a {\it tree} sub-graph of $G$ with $d_x=d_y=1$ and $d_z=2$ for any
vertex $z$ in $\t(x,y)$ distinct from $x$ and $y$.
We define the {\it distance}
between  $x$ and $y$ as 
$|x-y|= \min\{|\t(x,y)|: {\t(x, y)~{\rm path ~in~}G}\}$. Remark that
$|x-y|=1\Leftrightarrow \{x,y\}\in E$.

 Given $G=(V, E)$ connected and $R\subset V$, let $E|_R=\{\{x,y\}\in E: x\in R, y\in R\}$ 
and define the graph $G|_{R}=(R, E|_R)$. Note that $G|_{R}$ is a sub-graph of $G$. 
We call $G|_{R}$ {\it the restriction of $G$ to $R$}. We say 
that {\it $R\subset V$ is connected}
if $G|_{R}$ is connected.
For any non void $R\subset V$, we further denote by 
$\partial R$ the {\it external boundary} of $R$ which is the subset of 
$V\backslash R$ 
given by
$$
\partial R = \{y\in V\backslash R: \exists x\in  R: |x-y|=1\}\Eq(bounda)
$$
An {\it automorphism} of a graph $G=(V,E)$ is a bijective map
$\g:V\to V$ such that $\{x,y\}\in E \Rightarrow \{\g x, \g y\}\in E$.
\def\OO{{\cal O}}

 A graph $G=(V, E)$ is called {\it transitive} if, for any $x, y$
in $V$, it exists an automorphism $\g$ on $G$ which maps $x$ to $y$. The graph $G$
is called {\it quasi-transitive} if $V$ can be partitioned in finitely many sets 
$O_1, \dots O_s$ (vertex orbits) such that for $\{x, y\}\in O_i$ 
it exists an automorphism $\g$ on $G$ which maps $x$ to $y$ and this
holds
for all $i=1, \dots , s$. If $x\in O_i$ and $y\in O_i$ we say that
$x$ and $y$ are equivalent. 
Remark that a locally finite quasi-transitive graph is necessarily
bounded degree.

 Roughly speaking in a transitive graph  any
vertex of the graph is equivalent; in other words
$G$ ``looks the same'' by observers sitting in different vertices.  
In a quasi-transitive graph there is
a finite number of different type of vertices and $G$ ``looks the same''
by observers sitting in vertices of the same type.

 As a immediate example all periodic
lattices with the elementary cell made by one site (e.g. square lattice,
triangular lattice, hexagonal lattice, etc.) are transitive infinite graphs, while 
periodic
lattices with the elementary cell made by more than one site are
quasi-transitive infinite graphs.

Let $\mathbb{G}=(\mathbb{V}, \mathbb{E})$ be a connected infinite graph. 
$\mathbb{G}$ is said to be {\it amenable} if 
$$
\inf\left\{{|\partial W|\over|W|}: W\subset \mathbb{V}, ~0<|W|<+\i\right\}=0
$$

\ni A sequence $\{V_N\}_{N\in \mathbb{N}}$  of
finite sub-sets of $\mathbb{V}$ is called a {\it F\o lner sequence} if 
$$
\lim_{N\to \i}{ |\partial V_N|\over|V_N|}=0
\Eq(Folner)
$$
From now on  $\mathbb{G}=(\mathbb{V}, \mathbb{E})$ will denote 
a  connected locally finite infinite graph and $V_N\subset \VU$  a {\it finite} subset.

The  partition function of the {\it antiferromagnetic}
Potts model with $q$ colours on $\GI|_{V_N}$  {\it at zero temperature} 
can be rewritten (in a slightly different notation respect \equ(2.1.0)) as
$$
Z_{\GI|_{V_N}}(q)=\sum_{\sigma_{V_N}} 
\exp \left\{- \sum_{\{x,y\}\in {\rm P}_2(V_N)}J_{xy}\d_{\sigma_x \sigma_y} \right\} \Eq(2.4b)
$$
where
$$
 J_{xy}=\cases{ +\infty &if $|x-y|=1$ \cr\cr
0 & otherwise}\Eq(2.2b0)
$$
 We stress again that, due to assumption \equ(2.2b0) 
(i.e. antiferromagnetic interaction $+$ zero temperature), the function
$Z_{\GI|_{V_N}}(q)$
represents the number of ways that 
the  vertices $x\in V_N$ of $\GI|_{V_N}$ can be assigned
``colors'' from the set $\{1, 2, \dots , q\}$ in such way that adjacent vertices
always receive different colors. We also recall that the function 
$Z_{\GI|_{V_N}}(q)$ is called, in the graph theory language,
the {\it chromatic polynomial} of $\GI|_{V_N}$.

\vskip.2cm

\ni {\bf Definition 1}. {\it Let $\GI=(\VU,\mathbb{E})$ connected and locally finite infinite
graph and let $\{V_N\}_{N\in\mathbb{N}}$ be a F\o lner sequence of 
subsets of  $\VU$. Then we define, if it exists,  the
ground state specific entropy of the antiferromagnetic Potts model at zero temperature
on $\mathbb{G}$ as
$$
S_{\mathbb{G}}(q)= \lim_{N\to \infty} {1\over  |V_N|} \ln Z_{\GI|_{V_N}}(q)\Eq(red)
$$
We also define  the reduced
ground state degeneracy per site as 
$$
W_r(\mathbb{G}, q) = {1\over q} \lim_{N\to \i} \left[Z_{\GI|_{V_N}}(q)\right]^{1\over |V_N|}\Eq(Wr)
$$}
The ground state specific entropy $S_{\mathbb{G}}(q)$ and the reduced
ground state degeneracy $W_r(\mathbb{G}, q)$ are directly related by the identity
$$
S_{\mathbb{G}}(q)= \ln W_r(\mathbb{G}, q)+ \ln q\Eq(Ws)
$$

We can now state our main result.
\vskip.2cm
\ni {\bf Theorem 2}. {\it  Let $\mathbb{G}=(\VU,\mathbb{E}) $ a locally finite connected
quasi-transitive amenable infinite graph with  maximum degree $\D$, and
let $\{V_N\}_{N\in\mathbb{N}}$ a F\o lner sequence in $\GI$.
Then,  $W_r(\mathbb{G}, q)$  exists, is finite, is independent
on the choice of the sequence $\{V_N\}_{N\in\mathbb{N}}$, and
is analytic in the variable $1/q$ whenever
$|1/q|< {1/ 2e^3 \D}$ ($e$ being the basis of natural logarithm). }
\vskip.1cm

Again we stress that this result 
proves  the Schrock  conjecture in a considerably
stronger formulation, since
any regular lattice is a quasi-transitive amenable graph
but the class of quasi-transitive amenable graphs is
actually much wider than that of regular lattices. 

We remark also that the proof of analyticity of $W_r(\mathbb{G}, q)$
requires to prove the 
analyticity and boundness of the function ${|V_N|^{-1}} \ln Z_{\GI|_{V_N}}(q)$
for any finite graph $\GI|_{V_N}$ in a disk 
$|1/q|< {1/ C\D}$ {\it uniformly in the volume
$V_N$}, which is  a stronger statement than theorem 5.1 in  \cite{So}, claiming
that the zeros of the function $Z_{\GI|_{V_N}}(q)$ lie in the disk $|q|<C\D$ for any
$\GI|_{V_N}$ finite with maximum degree $\D$. 

\vskip.7cm
\ni\S3. {\bf Polymer expansion and analyticity}
\numsec=3\numfor=1
\vskip.3cm
\ni We first rewrite the  partition function
of the Potts model on a generic {\it finite} graph $G=(V,E)$ as a hard core
Polymer gas grand canonical partition function.
Without loss in generality, we will
assume in this section that  $G$  is a sub-graph of a 
bounded degree infinite graphs $\mathbb{G}=(\mathbb{V}, \mathbb{E})$ 
with  maximum degree  $\D$.
Denote by $\pi(V)$ the set
of all unordered partitions of $V$, i.e. an element of $\pi(V)$ is
an unordered n-ple $\{R_1,R_2,\dots ,R_n\}$, with $1\leq n\leq |V|$,
such that, 
for $i, j\in\I_n$, $R_i\subset V$, $R_i\neq \emptyset$, 
$R_i\cap R_j =\emptyset$, and $\cup_{i=1}^n R_i = V$. 
Then,  by writing the factor
$ \exp\{- \sum_{\{x,y\}\subset V} \d_{\sigma_x \sigma_y}J_{xy}\}$ in \equ(2.4b) as
$\prod_{\{x,y\}\subset V}[ (\exp\{- \d_{\sigma_x \sigma_y}J_{xy}\}-1)+1]$ and developing the
product (a standard Mayer expansion procedure, see e.g \cite{C}) we can rewrite 
the partition function on $G$  \equ(2.4b) as
$$
Z_{G}(q)=q^{|V|}\Xi_{G}(q)\Eq(2.5)
$$
where
$$
\Xi_{G}(q)=\sum_{n\ge 1}\sum_{\{R_1,\dots ,R_n\}\in \pi(V)}\r(R_1)\dots\r(R_n)\Eq(2.6)
$$
with 
\def\RR{{\bf R}}
$$
\r(R)=\cases{1 &if $|R|=1$\cr\cr
 q^{-|R|}\sum\limits_{\sigma_R\in \G_R}
\sum\limits_{E'\subset {\rm P}_2(R)\atop (R, E')\in {\cal G}_R}
\prod\limits_{\{x,y\}\in E'} [e^{-\d_{\sigma_x\sigma_y} J_{xy}}-1] 
&if $|R|\ge 2$ and $\GI|_R\in {\cal G}_R$ \cr\cr
0 &if $|R|\ge 2$ and $\GI|_R\notin {\cal G}_R$ 
}
\Eq(2.7)
$$
Observe that the sum  in l.h.s. of \equ(2.7) runs over all possible connected graphs with
vertex set $R$. 
The r.h.s. of \equ(2.6) can be written in a more compact way, by using the short
notations $$
\RR_n\equiv (R_1,\dots , R_n)
~~~~~~~~;
 ~~~~~~~{\r}(\RR_n)\equiv\r(R_1)\cdots\r(R_n)
$$
as
$$
\Xi_{G}(q)=1 + \sum_{n\ge 1}{1\over n!}
\sum_{\RR_n\in  [{\rm P_{\ge 2}}(V)]^n\atop R_i\cap R_j=\emptyset ~\forall \{i,j\}\subset \I_n}
\r(\RR_n)\Eq(XIG)
$$
\ni where $[{\rm P}_{\ge 2}(V)]^n$ denote the $n$-times Cartesian product of 
${\rm P}_{\ge 2}(V)$ (which, we recall, denotes the set of all finite subsets of $V$
with cardinality greater than 2).

It is also convenient to simplify the expression for
the activity
\equ(2.7) by performing the sum over $\sigma_R$. As a matter of fact
$$
q^{-|R|}\sum\limits_{\sigma_R\in \G_R}
\sum\limits_{E'\subset {\rm P}_2(R)\atop (R, E')\in {\cal G}_R}
\prod\limits_{\{x,y\}\in E'}  [e^{-\d_{\sigma_x\sigma_y} J_{xy}}-1]=
q^{-|R|}\sum\limits_{\sigma_R\in \G_R}
\sum\limits_{E'\subset {\rm P}_2(R)\atop (R, E')\in {\cal G}_R}
\prod\limits_{\{x,y\}\in E'}  \d_{\sigma_x \sigma_y}[e^{-J_{xy} }-1]
=
$$
$$=
q^{-|R|}\sum\limits_{E'\subset {\rm P}_2(R)\atop (R, E')\in {\cal G}_R}
\left[\sum\limits_{\sigma_R\in \G_R}\prod\limits_{\{x,y\}\in E'} \d_{\sigma_x \sigma_y}
\right]\prod\limits_{\{x,y\}\in E'} [e^{-J_{xy} }-1]
$$
But now, for any connected graph $(R, E')\in {\cal G}_R$
$$
\sum\limits_{\sigma_R\in \G_R}
\prod\limits_{\{x,y\}\in E'} \d_{\sigma_x \sigma_y}= q
$$
Hence we get, for $|R|>1$
$$
\r(R)=\cases{
q^{-(|R|-1)}
\sum\limits_{E'\subset {\rm P}_2(R)\atop (R, E')\in {\cal G}_R}
\prod\limits_{\{x,y\}\in E'} [e^{-J_{xy} }-1] & if $\GI|_R\in {\cal G}_R$ \cr\cr
0 & otherwise
}
\Eq(2.14)
$$
 By definitions \equ(2.14) or \equ(2.7), the polymer activity $\r(R)$ 
can be viewed as a real valued function 
defined on any  finite subset $R$ of $\mathbb{V}$. 
Of course this function depends on the ``topological structure'' of $\mathbb{G}$.
We remark that if $\g$ is an automorphism of $\mathbb{G}$, then \equ(2.14) clearly
implies that
$\r(\g R)=\r(R)$. In other words  the activity $\r(R)$ is invariant 
under automorphism of $\mathbb{G}$. 

The function $\Xi_{G}(q)$ is  the standard grand canonical partition function of
an {\it hard core polymer gas} in which the polymers are finite subsets $R\in V$ with
cardinality greater than $2$, with {\it activity } $\r(R)$, and submitted
to an {\it hard core} condition ($R_i\cap R_j=\emptyset$ for any pair $\{i,j\}\in \I_n$).

 Note that by \equ(2.5)and definitions \equ(Wr)-\equ(Ws) we have
$$
W_r(\mathbb{G}, q)=\exp\left\{\lim_{N\to \infty}{1\over |V_N|}\ln \Xi_{\GI|_{V_N}}(q)\right\}\Eq(yea)
$$

It is a well known fact in statistical mechanics  that 
the natural logarithm of $\Xi_{G}$   can be rewritten as formal series, called 
the {\it Mayer series} (see e.g. \cite{C}) as
\def\RR{{\bf R}}
$$
\ln\Xi_{G}(q)= \sum_{n= 1}^\i{1\over n!}
\sum_{\RR_n\in  [{\rm P_{\ge 2}}(V)]^n}
\phi^T(\RR_n)\r(\RR_n)\Eq(2.9)
$$
where 
$$
\phi^T(\RR_n)=\cases{\sum_{E'\subset E(\RR_n)\atop (\I_n,E')\in {\cal G}_n}
\prod_{\{i,j\}\in E'}(-1)^{|E'|} &if $G(\RR_n)\in {\cal G}_n$\cr\cr
0 &otherwise
}
\Eq(2.10)
$$
and $G(\RR_n)\equiv G(R_1,\dots ,R_n)$ defined at the beginning of section 2.
The reader should note that the summation in the l.h.s. of \equ(XIG) is
actually a {\it finite sum}. On the contrary, the summation 
in the l.h.s. of \equ(2.9) is  an {\it infinite series}.

We conclude this section showing two important technical lemmas concerning precisely the
convergence of the series \equ(2.10). 
In 
the proof of both lemma we will use a well known combinatorial  
inequality due to Rota \cite{R},
which states that if $G=(V,E)$ is a connected graph, i.e. $G\in {\cal G}_V$,  then
$$
\left|\sum_{E'\subset E:\atop (V, E')\in {\cal G}_V}(-1)^{|E'|}\right|\le
\sum_{E'\subset E:\atop (V, E')\in{\cal T}_V} 1= N_{{\cal T}_V}[G]\Eq(rota)
$$
where  $N_{{\cal T}_V}[G]$ is the number of tree 
graphs with vertex set $V$ which are sub-graphs of $G$.

\vskip.3cm
\ni {\bf Lemma 3}. {\it Let $\GI=(\VU, \mathbb{E})$ a bounded degree
infinite graph with maximum degree $\D$, and let, for any $R\in \VU$ such that
$|R|\ge 2$, the activity  $\r(R)$ be given as in \equ(2.14). Then, for any $n\ge 2$
$$
\sup_{x\in \mathbb{V}}\sum_{R\subset \VU:~ x\in R\atop |R|=n}|\r(R)|
\le\left[{e\D\over |q|}\right] ^{n-1} \Eq(bfk)
$$
}
\vskip.1cm
\ni {\bf Proof.} By definition
$$
\sup_{x\in \mathbb{V}}\sum_{R\in \VU:~ x\in R\atop |R|=n}|\r(R)|= 
|q|^{-(n-1)}\sup_{x\in \mathbb{V}}
\sum_{R\subset \mathbb{V}:~x\in R\atop |R|=n,~\GI|_R\in {\cal G}_R}\left|
\sum\limits_{E'\subset {\rm P}_2(R)\atop (R, E')\in {\cal G}_R}
\prod\limits_{\{x,y\}\in E'}  [e^{- J_{xy} }-1]\right|\Eq(fkk)
$$
Using thus the Rota inequality \equ(rota), recalling that
$\mathbb{E}|_R=\{\{x,y\}\in \mathbb{E}: x\in R, y\in R\}$, and observing that
$e^{- J_{xy} }-1=-1$ if $|x-y|=1$ and $e^{- J_{xy} }-1=0$ otherwise, we get
$$
\left|\sum\limits_{E'\subset {\rm P}_2(R)\atop (R, E')\in {\cal G}_R}
\prod\limits_{\{x,y\}\in E'}  [e^{-J_{xy} }-1]\right|=
\left|\sum\limits_{E'\subset \mathbb{E}|_R\atop (R,E')\in {\cal G}_R}
 (-1)^{|E'|}\right|
\le
\sum_{E'\subset \mathbb{E}|_R:\atop (R, E')\in {\cal T}_R} 1
=
\sum\limits_{E'\subset {\rm P}_2(R)\atop (R, E')\in {\cal T}_R}
\prod\limits_{\{x,y\}\in E'} \d_{|x-y|1}
$$
where $\d_{|x-y|1}=1$ if $|x-y|=1$ and $\d_{|x-y|1}=0$ otherwise. Hence
$$
\sup_{x\in \mathbb{V}}\sum_{R\subset \VU:~ x\in R\atop |R|=n,~\GI|_R\in {\cal G}_R}|\r(R)|
\le |q|^{-(n-1)}
\sup_{x\in \mathbb{V}}
\sum_{R\subset \VU:~ x\in R\atop |R|=n}
\sum\limits_{E'\subset {\rm P}_2(R)\atop (R, E')\in {\cal T}_R}
\prod\limits_{\{x,y\}\in E'}  \d_{|x-y|1}\le
$$
$$
\le {|q|^{-(n-1)}\over (n-1)!}
\sum\limits_{E'\subset {\rm P}_2(\I_n)\atop (\I_n, E')\in {\cal T}_n}
\left[\sup_{x\in \mathbb{V}}\sum_{x_1=x,\,(x_2,\dots , x_n)\in \VU^{n-1}\atop x_i\neq x_j~\forall \{i,j\}\in \I_n}
\prod\limits_{\{i,j\}\in E'}  \d_{|x_i-x_j|1}\right]
$$
It is now easy to check that, for any $E'\subset {\rm P}_2(\I_n)$ such that 
$(I_n, E')$ is a tree, 
it holds
$$
\sup_{x\in \mathbb{V}}\sum_{x_1=x,\,(x_2,\dots , x_n)\in \VU^{n-1}\atop x_i\neq x_j~\forall \{i,j\}\in \I_n}
\prod\limits_{\{i,j\}\in E'}  \d_{|x_i-x_j|1}\le {\D^{n-1}\over (n-1)!}
$$
and since, by Cayley formula, 
$\sum\limits_{E'\subset {\rm P}_2(\I_n)\atop (R, E')\in {\cal T}_n}1=n^{n-2}$, we get
$$
\sup_{x\in \mathbb{V}}\sum_{R\subset \VU:~ x\in R\atop |R|=n}|\r(R)|\le
\left({\D\over |q|}\right)^{n-1}{n^{n-2}\over (n-1)!}\le \left[{e\D\over |q|}\right] ^{n-1}
$$
$\square$

\vskip.2cm
To enunciate the second lemma  we need to introduce
a formal series more general that l.h.s.  of \equ(2.9).
Let thus $U\subset \VU$ finite and let $m$ a positive integer. We define 
$$
{\cal S}^m_{U}(\GI,q)=\sum_{n= 1}^\i{1\over n!}
\sum_{{\RR_n\in  [{\rm P}_{\geq 2}(\VU)]^n\atop  |\RR_n|\ge m,~  R_1\cap U\neq \emptyset}
}
\phi^T(\RR_n)\r(\RR_n)\Eq(Gen)
$$
 where $|\RR_n|=\sum_{i=1}^n|R_i|$ and recall that ${\rm P_{\ge 2}}(\VU)$ 
denotes the set of all finite subsets of $\VU$ with cardinality greater or equal
than 2 and
$[{\rm P_{\ge 2}}(\VU)]^n$ denote the $n$-times Cartesian product.
 We will now prove the following:

\vskip.3cm
\noindent{\bf Lemma 4}. {\it Let  $\mathbb{G}=(\mathbb{V}, \mathbb{E})$
 a locally finite infinite graph
with maximum degree $\D$. Let $U\subset \VU$ finite and  let $m$ a positive integer.
Then  ${\cal S}^m_{U}(\GI,q)$ defined in \equ(Gen)
exists  and is analytic as a function of $1/q$ in the disk \,
 $|2\D e^3/q|< 1$. Moreover it satisfies the following bound
$$
|{\cal S}^m_{U}(\GI)|\le  
|U|{1\over 1-\sqrt{2e^3|\D/q|}}
\left|2e^3{\D\over q}\right|^{m/2}
$$

}

\vskip.5truecm
\noindent
{\bf Proof}.

\noindent
We will prove the theorem
by showing directly that
the r.h.s. of \equ(Gen) converge absolutely when $|1/q|$ is sufficiently small.
Let us define
$$
|{\cal S}|^m_{U}(\GI)=\sum_{n= 1}^\i{1\over n!}
\sum_{{\RR_n\in  [{\rm P_{\ge 2}}(\VU)]^n\atop |\RR_n|\ge m,~  R_1\cap U\neq \emptyset}}
|\phi^T(\RR_n)\r(\RR_n)|
\Eq(10mod)
$$
then $|{\cal S}^m_{U}(\GI)|\le |{\cal S}|^m_{U}(\GI)$. We now bound $|{\cal S}|^m_{U}(\GI)$.
We have:
$$
|{\cal S}|^m_{U}(\GI)
\le \sum_{s=m}^{\i}\sum_{n= 1}^{[s/2]}{1\over n!}
\sum_{\RR_n\in  [{\rm P_{\ge 2}}(\VU)]^n \atop R_1\cap U\neq \emptyset,
~|\RR_n|=s}
|\phi^T(\RR_n)\r(\RR_n)|=
 \sum_{s=m}^{\i}\sum_{n= 1}^{[s/2]}{1\over n!}\sum_{{\bf k}_n\in \mathbb{N}^n:~k_i\ge 2\atop
k_1+\dots +k_n=s}B_n({\bf k}_n)
$$
where ${\bf k}_n\equiv (k_1, \dots ,k_n)$, $\mathbb{N}^n$ denotes the
$n$- times Cartesian product of $ \mathbb{N}$, 
$[s/2]=\max\{\ell\in \mathbb{N}: \ell\le {s/2}\}$, and 
$$
B_n({\bf k}_n)=
\sum_{{\RR_n\in  [{\rm P_{\ge 2}}(\VU)]^n\atop   R_1\cap U\neq \emptyset~~~~}
\atop|R_1|=k_1,\dots,\,|R_n|=k_n}
|\phi^T(\RR_n)\r(\RR_n)|
$$
recalling now \equ(2.10) and using again the 
Rota bound  \equ(rota) we get

$$
|\phi^T(\RR_n)| ~\cases{ 
\le N_{{\cal T}_n}[G(\RR_n)]~~~~~~ &if $G(\RR_n)\in {\cal G}_n$
\cr\cr
=0 &otherwise
}
$$
Hence
$$
B_n({\bf k}_n)\le \sum_{G\in {\cal G}_n }
N_{{\cal T}_n}[G]
\sum_{{\RR_n\in [P_{\ge 2}(\VU)]^n\atop   R_1\cap U\neq \emptyset,~
G(\RR_n)=G}
\atop|R_1|=k_1,\dots,|R_n|=k_n}
\left|
\r(\RR_n)\right|
\Eq(11)
$$
Observing now that 
$$
\sum_{G\in {\cal G}_n }
N_{{\cal T}_n}[G](\cdots)=
\sum_{\t\in {\cal T}_n}\sum_{G\in {\cal G}_n:~ G\supset \t}(\cdots)
$$
We can rewrite
$$
B_n({\bf k}_n)\le
\sum_{\t\in {\cal T}_n}B_n(\t,{\bf k}_n)\Eq(13b)
$$
where
$$
B_n(\t,{\bf k}_n)= 
\sum_{{\RR_n\in [P_{\ge 2}(\VU)]^n
\atop   R_1\cap U\neq \emptyset,~G(\RR_n)\supset \t}
\atop|R_1|=k_1,\dots,|R_n|=k_n}
\left|
\r(\RR_n)\right|
$$
Note now  that for any  non negative function $F(R)$ it holds
$$
\sum_{R\in \VU:\, R\cap R'\neq\emptyset\atop |R|=k}F(R)\leq
|R'|\sup_{x\in \mathbb{V}}\sum_{R\in \mathbb{V}\atop x\in R, ~|R|=k}F(R)\Eq(FR)
$$
Hence we can  now estimates
$B_n(\t,{\bf k}_n)$ for any fixed $\t$ by explicitly perform the sum
over polymers $\RR_n$ submitted to the constraint that
$g(\RR_n)\supset \t$, summing first over the 
``outermost polymers'', i.e. those polymers $R_i$ such that
$i$ is a vertex of degree 1 in $\t$, and  using  repetively the bounds \equ(FR).
Then one can easily check that

$$
B_n(\t,{\bf k}_n)
\le
|U|\sup_{x\in \mathbb{V}}\sum_{R_{1}\in \mathbb{V}\atop x\in R_1, ~|R_1|=k_1}
\cdots
\sup_{x\in \mathbb{V}}\sum_{R_n\in \mathbb{V}\atop x\in R_n ~|R_n|=k_n}
|\r(R_{1})|
|R_{1}|^{d_1}\prod_{i=2}^k\left[
|R_i|^{d_i -1}|\r(R_i)|\right]
\Eq(14)
$$
where $d_i$ is the degree of the vertex $i$ of $\t$. Recall that, 
for any tree $\t\in {\cal T}_n$,
it holds $1\le d_i\le n-1$ and $d_1+\dots +d_n=2n-2$. 
Now, by lemma 3, \equ(bfk),  we can bound
$$
B_n(\t,{\bf k}_n)
\le
|U|
\e^{k_1-1}
k_1^{d_1}\prod_{i=2}^k\left[
k_i^{d_i -1}\e^{k_n-1}\right]
\Eq(14b)
$$
where we have put for simplicity $\e= {e\D/ |q|}$.
Noting that estimates in l.h.s. of \equ(14)
depends only on the degrees $d_1, \dots , d_n$ of the vertices in $\t$,
we can now easily  sum over all connected tree
graphs in ${\cal T}_n$
and obtain
$$
B_n({\bf k}_n)\le
\sum_{\t\in{\cal  T}_n}B_n(\t,{\bf k}_n)=  
\sum_{{r_1, \dots ,r_n\atop r_1+\dots +r_n=2n-2}
\atop 1\le r_i\le n-1}
\sum_{\t\in {\cal T}_n\atop d_1=r_1, \dots , d_n=r_n}B_n(\t,{\bf k}_n)\le
$$
$$
\le {|U|}
\sum_{{r_1, \dots ,r_n\atop r_1+\dots r_n=2n-2}
\atop 1\le r_i\le n-1}
(n-2)!
{k_1}\prod_{i=1}^n\left[
{k_i^{r_i -1}\over (r_i -1)!} \e^{k_i-1}\right]
$$
where in the second line we used the bound \equ(14) and   Cayley formula
$$
\sum_{\t\in T_n\atop d_1, \dots d_n~{\rm fixed}} 1={(n-2)!\over\prod_{i=1}^{n}(d_i -1)!}\Eq(Cay)
$$
Now, recalling that $k_1+\dots +k_n=s$ and using the Newton multinomial formula, we get 
$$
B_n({\bf k}_n)\le  {|U|} k_1 s ^{n-2}\e^{s-n}\le {|V|}s^n\e^{s-n}
$$
thus, since $\sum_{k_1, \dots ,k_n:~k_i\ge 2\atop
k_1+\dots k_n=s}1\le 2^{s-n}$, we obtain
$$
|{\cal S}^m_{U}(\GI)|\le |U|\sum_{s=m}^{\i}
\sum_{n= 1}^{[s/2]}{s^n\over n!}\e^{s-n}\sum_{k_1, \dots ,k_n:~k_i\ge 2\atop
k_1+\dots k_n=s}1
\le 
|U|\sum_{s=m}^{\i}
\sum_{n= 1}^{[s/2]}{s^n\over n!}[2\e]^{s-n}
$$
The series above converge if $\e<{1\over 2e}$ and we get the bound
$$
|{\cal S}^m_{U}(\GI)|\le |U|\sum_{s=m}^{\i}
\sum_{n= 1}^{[s/2]}{s^n\over n!}[2\e]^{s-n}\le
\sum_{s=m}^{\i}[2\e]^{s-[s/2]}
\sum_{n= 1}^{\i}{s^n\over n!}\le
\sum_{s=m}^{\i}[2e^2\e]^{s/2}
\le
$$
$$
\le|U|{\left[2e^2\e\right]^{m/2}\over 1-e\sqrt{2\e}}
$$
provided
$$
2e^2\e<1
$$
Hence, recalling that $\e={e\D/|q|}$, the lemma is proved. $\square$

\vskip.1cm

The following corollary is now a trivial consequence of the two lemmas above.
\vskip.2cm
\ni {\bf Corollary 5}. {\it 
Let $G=(V,E)$ any finite connected sub-graph of an infinite connected bounded degree
graph $\mathbb{G}=(\mathbb{V},\mathbb{E})$ with maximum degree $\D$.
Then the function ${ |V|^{-1}}\log \Xi_{G}(q)$
is analytic in the variable $1/q$ for  $|1/q|< {1/ 2e^3 \D}$ and it admits the following
bound uniformly in $|V|$:
$$
\left|{1\over |V|}\log \Xi_{G}(q)\right|\le {1\over 1-\sqrt{2e^3|\D/q|}}
\left|2e^3{\D\over q}\right|\Eq(Sokall)
$$

}

\vskip.3cm

\ni {\bf Proof}. For any $G=(V,E)\subset \GI=(\VU,\mathbb{E})$ with $V$ finite,
by definition \equ(2.9) and \equ(Gen), 
it holds that $|\ln \Xi_{G}(q)|\le |{\cal S}|^2_{V}(\GI,q)$ and one can thus apply lemma 4.
$\square$

\vskip.7cm

\ni\S4. {\bf A graph theory lemma}
\numsec=4\numfor=1
\vskip.3cm

\ni {\bf Lemma 6}. {\it  Let $\mathbb{G}=(\mathbb{V},\mathbb{E})$ be
a locally finite quasi-transitive infinite graph and let $\{V_N\}_{N{\in}\mathbb N}$
be a F\o lner sequence of finite subsets of $\VU$. Then, for every vertex orbit $O\subset \VU$ of ${\rm Aut}(G)$,
there exists a non-zero finite limit
$$
\lim\limits_{N\to\infty}{{|O{\cap}V_N|}\over{|V_N|}}\Eq(limo)
$$
and it is independent on the choice of the sequence $\{V_N\}_{N{\in}\mathbb N}$.}
\vskip.1cm
\ni {\bf Proof}.
For a natural $r$ and a finite set $F{\subset}\VU$
denote by ${\rm B}_rF$ the set
$$
{\rm B}_r (F)= \{x{\in}\mathbb{V}:\exists y{\in}F\ |x-y|{\leqslant}r\}\Eq(g)
$$
Thus, for a single-point set $\{y\}$, ${\rm B}_r(\{y\})$ is 
the ball of radius $r$ centered at $y$. Moreover we have the bound
$$
|{\rm B}_r(F)|\leqslant|F|(1{+}\Delta{+}\dots{+}\Delta^r)\leqslant
\Delta^{r+1}|F|\Eq(l12)
$$
Let $O_1,\dots,O_s$ be the complete list of vertex orbits of ${\rm Aut}(\GI)$
in the set $\VU$ and let $A_0\subset \VU$ be a set with exactly one element in common
with every orbit. Denote by $d$ the diameter of $A_0$.
Consider the orbit ${\cal A}=\{gA_0:g{\in}{\rm Aut}(\GI)\}$ of $A_0$.
A set $A\subset \VU$  is therefore an element of ${\cal A}$ if
it exists a $g{\in}{\rm Aut}(\GI)\}$ such that $A=gA_0$.
For any set $U{\subset}V$ we denote
${\cal A}_U=\{A{\in}{\cal A}:A{\subset}U\}$.
Note that for any set $A\in {\cal A}$ and any vertex orbit $O$,
we have that $|A\cap O|=1$, hence
for a fixed vertex orbit ${O}_i$ we can define the function $\varphi_i$ as follows.
$$
\varphi_i: {\cal A}\to O_i: A\mapsto A\cap O_i
$$
The function $\varphi_i$ is a surjection and for $x\in O_i$ 
the number $k_i=|\varphi_i^{-1}(x)|$ is finite and does not depend on the choice of
$x\in O_i$.
For the sets
$$
V_N^-=V_N{\setminus}{\rm B}_d(\partial V_N),~~~~
V_N^+=V_N{\cup}{\rm B}_d(\partial V_N)\Eq(l13)
$$
we have $V_N^-{\subset}V_N{\subset}V_N^+$ and
$$
\varphi_i^{-1}(V_N^-{\cap}O_i)\subset
{\cal A}_{V_N}\subset\varphi_i^{-1}(V_N{\cap}O_i)\subset{\cal A}_{V_N^+}\Eq(l14)
$$
Indeed, suppose that $A{\in}\varphi_i^{-1}(V_N^-{\cap}O_i)$ and $A{\not\subset}V_N$.
Let $a_1{=}\varphi_i(A){\in}V_N^-{\subset}V_N$
and let $a_2{\in}A{\setminus}V_N$.
There exists a path $\t(a_1,a_2)$ in $\mathbb{G}$
of length ${\leqslant}d$. This chain must have at least one point in $\partial V_N$.
This implies $a_1{\in}{\rm B}_d(\partial V_N)$ contradicting the assumption
$a_1{\in}V_N^-$.
The second inclusion of \equ(l14) is obvious and the third one is true by the same
reason as the first one.
\equ(l14) implies
$$
k_i|V_N^-{\cap}O_i|\leqslant
|{\cal A}_{V_N}|\leqslant k_i|V_N{\cap}O_i|\leqslant|{\cal A}_{V_n^+}|\Eq(l15)
$$
and 
$$
|V_N^-{\cap}O_i|\leqslant\frac1{k_i}|{\cal A}_{V_N}|\leqslant
|V_N{\cap}O_i|\leqslant\frac1{k_i}|{\cal A}_{V_N^+}|\Eq(l15b)
$$
By taking sum over $i$ we get
$$
|V_N^-|\leqslant\alpha|{\cal A}_{V_N}|\leqslant
|V_N|\leqslant\alpha|{\cal A}_{V_N^+}|\Eq(l17)
$$
where $\alpha=\sum_{i=1}^s\frac1{k_i}$.
On the other hand 
${\cal A}_{V_N^+}{\setminus}{\cal A}_{V_N}\subset{\cal A}_{{\rm B}_d(\partial V_N)}$
and, by \equ(l12),
$$
|{\cal A}_{V_N^+}|\leqslant
|{\cal A}_{V_N}|+|{\cal A}_{{\rm B}_d(\partial V_N)}|\leqslant
|{\cal A}_{V_N}|+k\Delta^{d+1}|\partial V_N|\Eq(l18)
$$
where $k{=}{\rm max}\{k_i:i{\in}\{1,\dots,s\}\}$.
From the first inequality of \equ(l17) we have
$$
|{\cal A}_{V_N}|\geqslant
\frac1\alpha|V_N^-|\geqslant\frac1\alpha(|V_N|-|{\rm B}_d(\partial V_N)|)\geqslant
\frac1\alpha(|V_N|-\Delta^{d+1}|\partial V_N|)\Eq(l19)
$$
If $\frac{|\partial V_N|}{|V_N|}\leqslant\varepsilon$
then, by \equ(l18) and \equ(l19),
$$
1\leqslant\frac{|{\cal A}_{V_N^+}|}{|{\cal A}_{V_N}|}\leqslant
1+\frac{\Delta^{d+1}|\partial V_N|}{\frac1\alpha(|V_N|-\Delta^{d+1}|\partial V_N|)}
\leqslant1+\frac{\Delta^{d+1}\varepsilon}{\frac1\alpha(1-\Delta^{d+1}\varepsilon)}
$$
This proves that
$$
\lim_{N\to\infty}
\frac{|{\cal A}_{V_N^+}|}{|{\cal A}_{V_N}|}=1\Eq(l110)
$$
By \equ(l17) and \equ(l19) we also have
$$
\lim\limits_{N\to\infty}{|{\cal A}_{V_N}|\over |V_N|}={1\over\alpha}\Eq(l110b)
$$
Dividing \equ(l15b) by $|V_N|$
and using \equ(l110b) we obtain
$$
\lim_{N\to\infty}\frac{|O_i{\cap}V_N|}{|V_N|}={\frac1{k_i\alpha}}
$$ 
and 
the lemma is proved. $\square$

\vskip.7cm
\ni\S5. {\bf Potts model on infinite graphs: proof of theorem 2.}
\numsec=5\numfor=1
\vskip.3cm

 Let $\mathbb{G} =(\mathbb{V}, \mathbb{E})$ infinite bounded degree
and let $x\in \mathbb{V}$. Then we define
$$
f_\GI(x)= \sum_{n= 1}^\i{1\over n!}
\sum_{{\RR_n\in [{\rm P}_{\ge 2}(\VU)]^n\atop  x\in R_1}}
\phi^T(\RR_n){\r(\RR_n)\over|R_1|}\Eq(fi)
$$
We stress that, by construction, $f_\GI(x)$ is invariant under automorphism. 
I.e. if $x\in \mathbb{V}$ and $y\in \mathbb{V}$
are equivalent  (i.e. it exists $\g$ automorphism of $\mathbb{G}$ such that
$y=\g x$) then $f_{\GI}(x)= f_\GI(y)$.

 Given now a  {\it finite}  set $V_N\subset \VU$, 
we  define 
$$
F(V_N)={1\over |V_N|}\sum_{x\in V_N}f_\GI(x)\Eq(limm)
$$
The numbers $F(V_N)$ are actually functions of $q$.
As a trivial corollary of  lemma 4 we can state the following
\vskip.1cm

\ni {\bf Lemma 7}. {\it Let $\mathbb{G} =(\mathbb{V}, \mathbb{E})$ infinite bounded degree. 
Then for any $V_N\subset V$ finite, the functions $f_\GI(x)$ and $F(V_N)$ 
defined in \equ(fi) and \equ(limm) are analytic in the variable $1/q$ for  
$|1/q|< {1/ 2e^3 \D}$  and bounded by   $\left|2e^3{\D/ q}\right|/(1-\sqrt{2e^3|\D/q|)}$ 
uniformly in $N$.}
\vskip.1cm

\ni{\bf Proof}. Comparing l.h.s. of  \equ(Gen) with l.h.s. of \equ(fi) we have 
that $|f_\GI(x)|\le |{\cal S}|_{\{x\}}^2(\GI)$, hence one can again use lemma 4 
and get immediately
the proof.  
$\square$.

\vskip.1cm
\ni From lemma 6 and lemma 7 it follows: 
\vskip.2cm
\ni {\bf Proposition 8}. {\it 
Let $\mathbb{G}=(\mathbb{V},\mathbb{E})$ be
a locally finite quasi-transitive infinite graph and let $\{V_N\}_{N{\in}\mathbb N}$
be a sequence of finite subsets of $\VU$ 
such that ${|\partial V_N|}/{|V_N|\to 0}$ as ${N\to\infty}$.
Let $\D$ be the
maximum degree of $\GI$,
then the limits
$$
\lim_{N\to \infty} F(V_N) \doteq F_{\mathbb{G}}(q)\Eq(liF)
$$
exists, is finite, is independent on the sequence $\{V_N\}_{N{\in}\mathbb N}$, and is analytic 
as a function of $1/q$  for  $|1/q|< {1/ 2e^3 \D}$.

}

\vskip.2cm
\ni{\bf Proof}. If the limit \equ(liF) exists, 
then by lemma 7
it is clearly bounded by $|2e^3{\D/ q}|/( 1-\sqrt{2e^3|\D/q|})$ 
and it
analytic in $1/q$ for  $|1/q|< {1/ 2e^3 \D}$ . 
To prove the existence of the limit \equ(liF) we proceed
as follows.

 Since $\mathbb{G}$ is quasi-transitive then $\mathbb{V}$ can be  partitioned
into orbits ${O}_1$, \dots , ${O}_s$  of ${\rm Aut} (\GI)$
such that for two any vertices $x, y$
in the same orbit ${ O}_i$ there is an automorphism of $\mathbb{G}$ which maps $x$ to $y$.
Hence for such a pair we  have $f_\GI(x)=f_\GI(y)$ and we can conclude that
$f_\GI(x)$ has value in a finite set $\{f_1, \dots , f_s\}$ with
$f_i=f_\GI(x)$ where $x$ is any vertex  $x\in {\cal O}_i$.

 Thus for any finite connected ${V_N}$ and any $j\in \{1, 2, \dots , s\}$ we have
$$
{1\over |V_N|}\sum_{x\in V_N}f_\GI(x)= 
\left[{|V_N\cap O_1|\over |V_N|} f_1+ \dots {|V_N\cap O_s|\over |V_N|} f_s\right]
$$
hence
$$
\lim_{N\to \i} F(V_N)=
=f_1\lim_{N\to \infty} {|V_N\cap O_1|\over |V_N|}+\dots + 
f_s\lim_{N\to \infty} {|V_N\cap O_s|\over |V_N|}
$$
and
by lemma 6 the limit above exists. $\square$

We are at last in the position to prove the main results of the paper, namely
the theorem 2 enunciated at the end of section 2.

\vskip.2cm

\ni {\bf Proof of theorem 2}.
We will prove that $
\lim_{N\to \i}{ |V_N|^{-1}}\log \Xi_{\GI|_{V_N}}(q)= F_{\mathbb{G}}(q)$
where $ F_{\mathbb{G}}(q)$ is the function defined in \equ(liF) and
then use definition \equ(yea).

$$
\log \Xi_{\GI|_{V_N}}- \sum_{x\in V_N}f_\GI(x)
=
\sum_{n=1}^\i{1\over n!}\left[
\sum_{\RR_n\in [{\rm P}_{\ge 2}(V_N)]^n}
\phi^T(\RR_n)\r(\RR_n)-
 \sum_{x\in V_N} 
\sum_{{\RR_n\in [{\rm P}_{\ge 2}(\VU)]^n\atop  x\in R_1~~~~}}
\phi^T(\RR_n){\r(\RR_n)\over |R_1|}\right]
$$
Now note that
$$
\sum_{{\RR_n\in [{\rm P_{\ge 2}}(\VU)]^n\atop x\in R_1}}(\cdots )=
\sum_{{\RR_n\in [{\rm P}_{\ge 2}(V_N)]^n\atop x\in R_1}}(\cdots )+
\sum_{{\RR_n\in [{\rm P}_{\ge 2}(\VU)]^n\atop x\in R_1}
\atop\exists R_i: ~
R_i\not\subset V_N}(\cdots )
$$
moreover 
$$\sum_{x\in V_N}\sum_{R_1\in V_N\atop x\in R_1}(\cdots)=
\sum_{R_1\in V_N}|R_1|(\cdots)~~~,~~~
\sum_{x\in V_N}\sum_{R_1\in \VU\atop x\in R_1}(\cdots)=
\sum_{R_1\in \VU}|R_1\cap V_N|(\cdots)
$$
\ni hence, using also that ${|R_1\cap V_N|/ |R_1|}\le 1$ we get
$$\Big|\log \Xi_{\GI|_{V_N}}- \sum_{x\in V_N}f_\GI(x)\Big|\le
\sum_{n=1}^\i{1\over n!}
\sum_{{\RR_n\in [{\rm P}_{\ge 2}(\VU)]^n\atop  R_1\cap V_N\neq \emptyset}
\atop\exists R_i: ~
R_i\not\subset V_N }
\left|\phi^T(\RR_n)\r(\RR_n)\right|
$$
Let  now choose $p>\ln \D$ and define
$$
m_N^p ={1\over p}\ln \left[{|V_N|\over |\partial V_N|}\right]\Eq(Bns)
$$
remark that, since by the hypothesis the sequence $ V_N$ is F\o lner and hence
\equ(Folner) holds, then  $\lim_{N\to \infty}m_N^p=\infty$, for any integer $p$.
We now can rewrite
$$
\sum_{{\RR_n\in [{\rm P}_{\ge 2}(\VU)]^n\atop R_1\cap V_N\neq \emptyset}
\atop\exists R_i: ~
R_i\not\subset V_N }(\cdots)=
\sum_{{\RR_n\in [{\rm P}_{\ge 2}(\VU)]^n\atop  R_1\cap V_N\neq \emptyset,~
|\RR_n|\ge m_N^p}\atop\exists R_i: ~
R_i\not\subset V_N , ~}(\cdots)+
\sum_{{\RR_n\in [{\rm P}_{\ge 2}(\VU)]^n\atop R_1\cap V_N\neq \emptyset,~|\RR_n|< m_N^p}
\atop\exists R_i: ~
R_i\not\subset V_N, }(\cdots)
$$

 Hence
$$
\left|\log \Xi_{\GI|_{V_N}}- \sum_{x\in V_N}f_\GI(x)\right|\le
\sum_{n=1}^\i{1\over n!}\sum_{{\RR_n\in [{\rm P}_{\ge 2}(\VU)]^n\atop  R_1\cap V_N\neq \emptyset,~
|\RR_n|\ge m_N^p}\atop\exists R_i: ~
R_i\not\subset V_N , ~}
 \left|\phi^T(\RR_n)\r(\RR_n)\right|+
$$
$$
+
\sum_{n=1}^\i{1\over n!}\sum_{{\RR_n\in [{\rm P}_{\ge 2}(\VU)]^n\atop R_1\cap V_N\neq \emptyset,~|\RR_n|< m_N^p}
\atop\exists R_i: ~
R_i\not\subset V_N, }
\left|\phi^T(\RR_n)\r(\RR_n)\right|\Eq(d)
$$
but, concerning the first sum, recalling definition \equ(10mod), we
have
$$
\sum_{n=1}^\i{1\over n!}\sum_{{\RR_n\in [{\rm P}_{\ge 2}(\VU)]^n\atop  R_1\cap V_N\neq \emptyset,~
|\RR_n|\ge m_N^p}\atop\exists R_i: ~
R_i\not\subset V_N , ~}
 \left|\phi^T(\RR_n)\r(\RR_n)\right|\le 
|{\cal S}|^{m_N^p}_{V_N}(\GI,q)\le
{\rm Const.}|V_N|\e^{m_N^p/2}
$$
where $\e= q/2e^3\D<1$ by hypothesis.
which, divided by $|V_N|$,  converge to zero as ${N}\to \i$ because
by hypothesis $m_N^p\to \i$ as $N\to \i$.

 On the other hand, recalling that due to the factor $\phi^T(\RR_n)$
the sets $R_i$ must be pair-wise
connected, we have that $|\cup_i R_i|<\sum_i|R_i|$. So, since $|\cup_i R_i|< m_N^p$ 
and at least one among $R_i$
intersects $\partial V_N$,
this means that all polymers $R_i$ must lie in the set
$$
{\rm B}_{m_N^p}(\partial V_N)=\{x\in \mathbb{V}:  \exists v\in \partial V_N : |x-v| \le m_N^p\}
$$
Recalling \equ(g) we have
$$
|{\rm B}_{m_N^p}(\partial V_N)|\le |\partial V_N| \D^{m_N^p+1}
$$
Hence we have, again recalling \equ(10mod),  that second sum in l.h.s. of \equ(d) is 
bounded by
\vskip.2cm
$$
\sum_{n=1}^\i{1\over n!}
\sum_{{\RR_n\in [{\rm P}_{\ge 2}(\VU)]^n\atop R_1\cap V_N\neq \emptyset,~|\RR_n|< m_N^p}
\atop\exists R_i: ~
R_i\not\subset V_N, }
\left|\phi^T(\RR_n)\r(\RR_n)\right|
\le |{\cal S}|^{2}_{{\rm B}_{m_N^p}(\partial V_N)}(\GI,q)\le
$$
$$
\le {\rm Cost.}|{\rm B}_{m_N^p}(\partial V_N)|\e\le
{\rm Const.}  \D|\partial V_N| \D^{m_N^p}\e
$$

\ni Thus  recalling definition
\equ(Bns), we have
$$
\left|{1\over |V_N|}\log \Xi_{\GI|_{V_N}}- 
{1\over |V_N|}\sum_{x\in V_N}f_\GI(x)\right|=
\left|{1\over |V_N|}\log \Xi_{\GI|_{V_N}}- 
F_{\mathbb{G}}(q)\right|
\le 
$$
$$
\le {\rm Const.}  \left[{|\partial V_N|\over |V_N|}\right]^{|\ln\e|\over p}
+
{\rm Const.} \e \left[{|\partial V_N|\over  |V_N|}\right]^{1-{\ln\D\over p}}
$$
Since by hypothesis $|\partial V_N|/ |V_N|\to 0$ as $N\to \i$, we conclude that the quantity above
is as small as we please for $N$ large enough. This ends the proof of the theorem. $\square$
\vskip.3cm

\vglue.5truecm
 {\bf Acknowledgements} 

 This work was supported by Ministero dell'Universita' e della
Ricerca Scientifica e Tecnologica (Italy) and Conselho Nacional de
Desenvolvimento Científico e Tecnológico  - CNPq, a Brazilian
Governmental agency promoting Scientific and technology
development (grant n. 460102/00-1).

 We thank Prof. M\'ario Jorge Dias Carneiro for useful 
discussions and Prof. Bojan Mohar for a valuable suggestion via e-mail.
\vglue.5truecm


\end{document}